\documentclass[aps,twocolumn]{revtex4}

\usepackage{graphicx}
\usepackage{color}
\usepackage{topcapt}
\usepackage{booktabs}
\usepackage{amsmath, amsthm, amssymb}

\begin{document}

\bibliographystyle{apsrev}

\title{Characterizing Spinning Black Hole Binaries in Eccentric Orbits with LISA }
\author{\surname{Joey} Shapiro Key and {Neil} J. Cornish}

\affiliation{Department of Physics, 
  Montana State University, Bozeman, MT 59717}
\date{\today}

\begin{abstract}

The Laser Interferometer Space Antenna (LISA) is designed to detect gravitational wave signals from
astrophysical sources, including those from coalescing binary systems of compact objects such as black holes.
Colliding galaxies have central black holes that sink to the center of the merged galaxy and begin to orbit
one another and emit gravitational waves.  Some galaxy evolution models predict that the binary black hole
system will enter the LISA band with significant orbital eccentricity, while other models suggest that the
orbits will already have circularized. Using a full seventeen parameter waveform model that includes the effects of
orbital eccentricity, spin precession and higher harmonics, we investigate how well the source parameters can be
inferred from simulated LISA data. Defining the reference eccentricity as the value one year before merger,
we find that for typical LISA sources, it will be possible to measure the eccentricity to an accuracy of
parts in a thousand. The accuracy with which the eccentricity can be measured depends only very weakly on
the eccentricity, making it possible to distinguish circular orbits from those with very small eccentricities.
LISA measurements of the orbital eccentricity can help constraints theories of galaxy mergers in the early
universe. Failing to account for the eccentricity in the waveform modeling can lead to a loss of signal power
and bias the estimation of parameters such as the black hole masses and spins.
\end{abstract}

\maketitle

\section{Introduction}

\begin{figure}[htbp]
   \centering
         \includegraphics[width=3in]{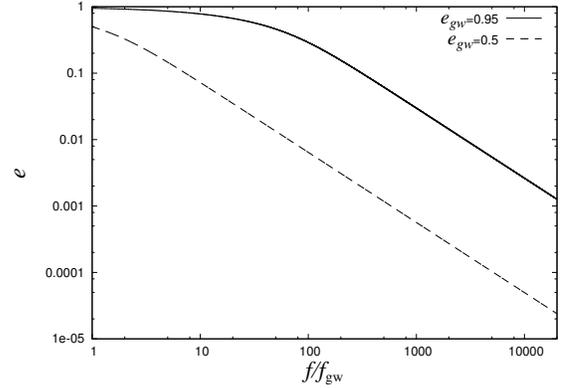} 
   \caption{Gravitational wave driven eccentricity evolution as a function of orbital frequency.}
   \label{fig:etrack}
\end{figure}

Binary systems of compact objects will be ubiquitous sources for the Laser Interferometer Space Antenna (LISA)~\cite{Micic:2007vd,Sesana:2007sh}.  Observations have shown that today there are massive black holes in the center of nearly all galaxies ~\cite{Kormendy:1995er,Magorrian:1997hw,Narayan:2005oq}.  When galaxies collide their central black holes sink to the center of the merged galaxy and begin to orbit one another, losing energy and angular momentum in the form of gravitational waves~\cite{Begelman:1980vb}. Gravitational wave (GW) emission rapidly erases any
initial eccentricity~\cite{Peters:63,Peters:1964en}, so it has long been thought that eccentricity could be ignored when
modeling the signals from massive black hole binaries~\cite{Krolak:1987vp}. More recently, however,
it has been shown that the mechanisms that may harden the binary to the point where the gravitational
wave emission takes over all tend to drive up the
eccentricity~\cite{Colpi:1999cm,Gultekin:2005fd,Dotti:2006ae,Berczik:2005ff,Armitage:2005xq,Aarseth:2002ie,AmaroSeoane:2006py, AmaroSeoane:2007aw,AmaroSeoane:2009yr,AmaroSeoane:2009cg}. The question then is whether significant
eccentricity survives until the final year or so before merger. Figure~\ref{fig:etrack} shows the eccentricity
evolution as a function of orbital frequency for systems that enter the gravitational wave dominated
evolution stage at frequency $f_{\rm gw}$ with eccentricities of $e_{\rm gw} = 0.95$ and $e_{\rm gw} = 0.5$.
The tracks are computed using the leading-order Peters and Matthews~\cite{Peters:1964en} evolution
equations. These equations predict that once the eccentricity drops below $e\sim 0.3$, it decays as
$e \sim f^{-19/18}$ - in other words, roughly a factor of ten in eccentricity is lost per decade
of frequency. The rate of decay is slower for systems with very high eccentricities, allowing them
to maintain significant eccentricity for longer. For typical LISA black hole binaries,
the GW decay time drops below the Hubble time when the systems are 3 to 5 decades in frequency from
the LISA band, so unless $e_{\rm gw}$ is very close to unity, a purely GW driven orbital evolution
results in nearly circular orbits in the LISA band. On the other hand, if the hardening
mechanism ({\it e.g.} gas dynamics or stellar scattering) continues to dominate the dynamics
until the system is close to the LISA band, then interesting eccentricities can be maintained.
In a recent study~\cite{Sesana:2010qb}, Sesana has shown that stellar scattering produces LISA sources
with eccentricities in the range $e_0 = 10^{-3} \rightarrow 0.2$. Moreover, as shown in Figure 8 of
Ref.~\cite{Sesana:2010qb}, the distribution of eccentricities depends on the component masses in a particular
way, so measuring this distribution can help constrain black hole merger models. While the distribution of
component masses and spins as a function of luminosity distance will likely play a more important role in
studying galaxy - black hole co-evolution~\cite{Plowman:2009rp,Sesana:2010wy,Plowman:2010fc}, the eccentricity
distribution may provide useful additional constraints. These considerations suggest that it is desirable to include the
effects of eccentricity in the gravitational waveform, bringing the total number of parameters needed to describe a black hole inspiral to seventeen: the redshifted black hole masses $\left(m_1,m_2\right)$, the distance to the source $\left(D_L\right)$, the initial radial eccentricity and semi-major axis $\left(e_0,a_0\right)$, the dimensionless spin parameters for the two black holes $\left(\chi_1,\chi_2\right)$, the source sky location $\left(\cos{\theta},\phi\right)$, the initial orientation of the angular momentum and spin vectors $\left(\cos{\theta_L},\phi_L\right),\left(\cos{\theta_{S_1}},\phi_{S_1}\right),\left(\cos{\theta_{S_2}},\phi_{S_2}\right)$, and two initial phase parameters $\left(n_0,\phi_0\right)$.  Failing to include the eccentricity in the waveform model will lead to a loss of signal-to-noise~\cite{Martel:1999kb,Brown:2009ng} and to biases in the recovery of the other parameters~\cite{Cutler:2007mi,Porter:2010mb}. 

The instantaneous gravitational waveforms describing the inspiral of a black hole binary with arbitrary spins, masses, and orbital eccentricity were calculated by Kidder~\cite{Kidder:1995zr} to second post-Newtonian order (2\,PN, or order $v^4/c^4$ in the relative velocity), and extended to 2.5\,PN order by Faye, Blanchet and Buonanno~\cite{Faye:2006gx,Blanchet:2006gy}.  Maj\'{a}r and Vas\'{u}th ~\cite{Majar:2008zz} later introduced a convenient framework for expressing the waveforms in a form amenable to producing detection templates.  Together with a solution to the conservative equations of motion, adiabatically advanced with the angular momentum and energy dissipation equations~\cite{Kidder:1995zr}, we can build the time dependent gravitational waveforms for a general binary black hole system with the full seventeen parameters necessary to describe the system.  LISA observations of binary black hole inspirals can be compared to these waveforms to produce posterior distributions for the model parameters. These observations will allow us to constrain galaxy merger scenarios~\cite{Plowman:2009rp,Sesana:2010wy,Plowman:2010fc}, and allow us to test General Relativity in the dynamical
strong field regime~\cite{Yunes:2009ke}. Left unaccounted for, the effects of orbital eccentricity may be mistaken
for a departure from General Relativity (in particular, even small eccentricities that produce negligible power
in higher harmonics can lead to easily detectable changes in the phase evolution of the signal).

Black hole binary systems in eccentric orbits may also be detected by ground based gravitational wave detectors such as the Laser Interferometer Gravitational wave Observatory (LIGO) and Virgo ~\cite{Wen:2003ec}.  Some models even predict that inspiral signals may enter the LIGO band with $e>0.9$ and that eccentric templates will be necessary to detect such sources ~\cite{OLeary:2008kx}.  Current LIGO data analysis uses circular templates and may need to be generalized to include eccentricity.

We have previously described a method for combining the instantaneous gravitational waveforms for eccentric binary systems with a post-Keplerian solution to the equations of motion that is adiabatically advanced using the the orbit-averaged dissipation and spin precession equations to build ready-to-use gravitational waveforms for the general case of a spinning black hole binary system in an eccentric orbit~\cite{Cornish:2010rz}.  Here we present the results of a parameter estimation study for spinning eccentric binary black hole sources for the LISA mission.  This  is an extension of the Lang and Hughes LISA parameter estimation study of spinning binary black holes in circular orbits~\cite{Lang:1900bz} and is the first to include the full seventeen parameters that describe a general black hole binary inspiral.
	
\section{Waveform Model}

The equations needed to numerically calculate time dependent gravitational waveforms for binary black hole
systems have been computed to 2.5PN order in the amplitude and phase~\cite{Kidder:1995zr,Faye:2006gx,Blanchet:2006gy,Majar:2008zz}. The resulting system of equations could be numerically evolved to produce waveforms
for our study, but the computational cost of resolving the motion of the black holes on the orbital timescale
is too large for the parameter estimation study we wish to perform. Taking advantage of the separation of
time scales in the waveform model we have developed an efficient waveform generator at 1.5 PN order in
the amplitude and phase~\cite{Cornish:2010rz}. These waveforms include the effects of periastron precession,
the precession of the orbital plane due to spin-orbit coupling, and higher harmonics from the higher order
mass and current multipole moments. Our waveform model does not include the effects of spin-spin coupling
which enter at 2PN order. In future work we plan to extend the waveform model and
our parameter estimation study to higher Post-Newtonian order.

The fastest time scale for the system is the orbital time scale, which to leading order is given by Kepler's law:
\begin{equation}
T_{\rm orb}\sim 2\pi a^{\frac{3}{2}} M^{-\frac{1}{2}} \, 
\end{equation}
where $M = m_1 + m_2$ is the total mass of the system and $a$ is the semi-major axis. Periastron advance
enters at 1\,PN order, with a fractional advance per orbit of $k= 3 M^2\mu^2/L^2$, where
$\mu = m_1 m_2/M$ is the reduced mass of the system, $L \simeq \mu \sqrt{a M (1-e^2)}$ is the
orbital angular momentum and $e$ is the eccentricity of the orbit. The ratio of the periastron precession timescale
to the orbital timescale is given by
\begin{equation}
\frac{T_{\rm peri}}{T_{\rm orb}} = \frac{2\pi}{k} \simeq \frac{a}{M} \, .
\end{equation}

The orbital angular momentum $\bf L$ precesses with angular velocity
${\boldsymbol \omega}_{\rm prec}= {\bf S}_{\rm eff}/{r^3}$~\cite{Cornish:2010rz}, where
\begin{equation}\label{Seff}
{\bf S_{\rm eff}} =  2 \left( 1 + \frac{3 m_2}{4 m_1} \right){\bf S}_1
+2 \left( 1 + \frac{3 m_1}{4 m_2} \right) {\bf S}_2
\end{equation}
The effective spin vector has magnitude $S_{\rm eff}\sim M^2$ since the magnitude of the individual spins is given by $S_i=\chi_i m^2_i$ where the dimensionless spin parameter $0 \leq \chi_i \leq 1$.  The ratio of the precession
time scale and the orbital timescale is given by
\begin{equation}
\frac{T_{\rm prec}}{T_{\rm orb}}\sim\left(\frac{a}{M}\right)^{\frac{3}{2}} \, .
\end{equation}
We see that the precession of the orbital plane enters at 1.5\,PN order relative to the orbital time scale.
The precession of the orbital angular momentum due to spin-orbit coupling results in a modulation
of the amplitude of the gravitational waveform at the solar system barycenter.


\begin{figure}[htbp]
   \centering
         \includegraphics[width=3in]{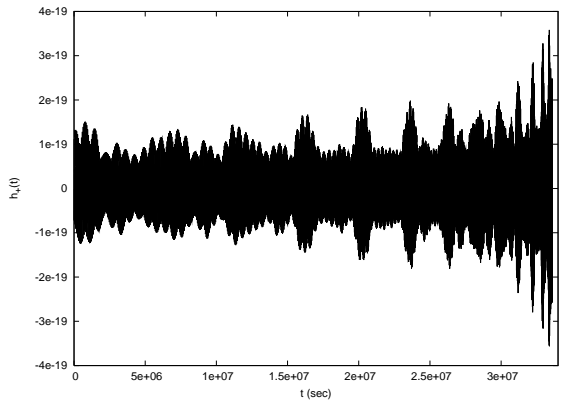} 
   \caption{The time dependent gravitational waveform $h_+(t)$ at the solar system barycenter for
a binary black hole system with $e_0=0.3$ (Source 1 in Table~\ref{tab:sourceP}) during the
final year before merger.}
   \label{fig:hyear}
\end{figure}

The rate at which the binary black hole system loses energy and angular momentum due to gravitational wave
emission defines the decay time scale
\begin{equation}
T_{\rm decay}\sim\frac{E}{\dot{E}}\sim\frac{L}{\dot{L}}
\end{equation}
and the ratio
\begin{equation}
\frac{T_{\rm decay}}{T_{\rm orb}}\sim\frac{M}{\mu}\left(\frac{a}{M}\right)^{\frac{5}{2}} \, .
\end{equation}
The loss of energy and angular momentum results in the decay of the semi-major axis and the radial
eccentricity of the system. Orbit averaged expression for these decay rates can be found in
Ref.~\cite{Cornish:2010rz}.  The effects of spin-orbit induced precession of the orbital plane
and the overall sweep up in frequency and amplitude as the system inspirals
over the course of a year are apparent in Figure~\ref{fig:hyear} for a system with $e_0=0.3$ and 
other source parameters given in Table~\ref{tab:sourceP} for Source 1.

At times well before the merger of the system $M/a$ is small and we find that
\begin{equation}
T_{\rm decay} > T_{\rm prec} > T_{\rm peri} > T_{\rm orb} \,.
\end{equation}
We take advantage of this separation of the relevant time scales to make our waveform calculations more
efficient.  Since the decay and precession timescales are much longer than the orbital timescale we can start with a solution to the orbital equations of motion that neglects dissipation.  This allows us to use Nyquist sampling of just a couple of samples per orbit for the quantities that depend on the dissipation and precession equations and saves considerable computational cost.  Considering only times early in the evolution of the system we have a clean separation of time scales and the different processes can be treated differently in the calculations.  This simplifies the problem by allowing an adiabatic treatment where the dissipation is assumed to be small over the course of an orbit and the eccentricity and semi-major axis are treated as constant while calculating an individual orbit.

The approximations involved in exploiting the separation of timescales introduce small errors in the waveforms.
The largest of these comes from using the orbit-averaged spin precession equations, which neglects small periodic
changes in the spin orientations that occur on the orbital timescale, but these changes are effectively 2.5 PN
order contributions to the higher harmonics. The averaging also introduces small errors in the long term secular
evolution that scale as the ratio of the averaging time scale (in our case the orbital timescale) and the
timescale of the terms being averaged (such as the spin precession timescale). These errors are multiplicative,
and so represent higher PN order terms that can be discarded at 1.5 PN order.

As the system approaches merger the various time scales become comparable and our
waveform model breaks down.  We adopt the termination condition $2 \pi M f_{\rm orb}=0.01$, which corresponds
to an expansion parameter $M/a \approx0.05$. Denoting the time when this condition is met as $t_s$, and
the orbital frequency at this time as $f_s$, we taper the waveform smoothly to zero by multiplying the
waveform with a half-Hann filter:
\begin{equation}
w(t)=
\begin{cases}
1, & \mbox{if } t \leq t_H \\
\cos^2\left(\pi(t-t_H)f_s/3 \right) , & \mbox{if } t > t_H
\end{cases}
\end{equation}
with $t_H=t_s - 3/(2 f_s)$.

Our termination condition is conservative in terms of the signal to noise ratio (SNR) LISA will be able to extract from this type of source.  Most of the SNR comes from times near merger, so the extension of the validity of the waveform closer to merger results in a big increase in SNR.  Our study here is thus a pessimistic estimate of how well LISA will be able to determine the various source parameters.  In the case of the radial eccentricity parameter however, the circularization of the waveform toward merger means that most of the eccentricity information is encoded at times well before merger.  While increased SNR would improve the determination of the other source parameters, we find that the eccentricity is not highly correlated with the other parameters (see Figure~\ref{fig:ecchist}).  Our choice for when to truncate the waveform thus should not have much of an effect on our study of how well LISA will be able to determine the eccentricity of black hole binary systems.

We have tested our waveform generator in various limits against other codes. In the circular limit, and with
dissipation turned off, we found precise agreement with the 1.5 PN limit of the spinning black hole code
developed by Cornish, Hughes, Lang and Nissanke that is described in Ref.~\cite{Arun:2008zn}. We do not
expect, and nor do we find, precise agreement when dissipation is included.  This is because our
eccentric waveform generator evolves the semi-major axis, while the circular orbit code evolves the
orbital frequency, which leads to numerical differences at 2 PN order. In the 0-PN limit
(hence no spin effects and no higher harmonics)
we find precise agreement with the Peters and Matthews waveforms~\cite{Peters:63}.

\section{LISA Response}\label{response}

We simulate the LISA response to a gravitational wave signal plus instrument noise and confusion noise due to galactic binary sources of gravitational waves.  We adopt the standard ecliptic coordinate system with origin at the barycenter.  The individual data streams from the six LISA phase meters can be combined to cancel out the laser phase noise and form Time Delay Interferometry (TDI) variables~\cite{Estabrook:2000ef}. The Michelson style TDI variables $\{X,Y,Z\}$ can be used to construct three noise orthogonal data streams that are similar to the $\{A,E,T\}$ variables described in Ref.~\cite{Prince:2002bx}.   

\begin{figure}[htbp]
   \centering
   \includegraphics[width=3in]{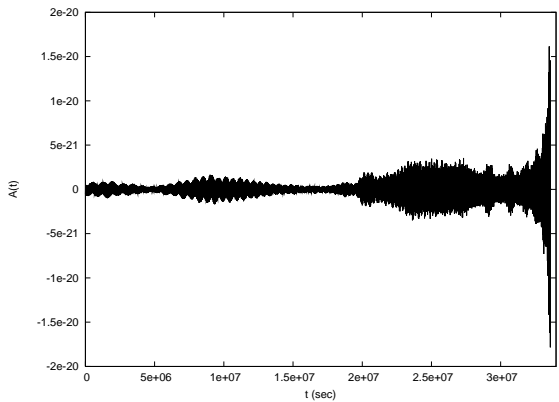}
   \caption{The $A(t)$ channel response to the spinning binary black hole system shown in
Figure~\ref{fig:hyear}.  The amplitude modulation is due to a combination of the
antenna pattern sweeping around as LISA orbits the Sun, and the spin induced precession of the
orbital plane. The overall gravitational wave amplitude grows as the system spirals
in and nears merger.}
   \label{fig:Achannel}
\end{figure}

Processing the gravitational waveform through the LISA response function imparts additional amplitude and
frequency modulations on the one year timescale of the LISA orbits. These effects can be seen in
Figure~\ref{fig:Achannel}, which shows the $A$-channel response to the Barycenter signal shown previously
in Figure~\ref{fig:hyear}.

The one-sided noise spectral density of the detector in the A and E channels is given by~\cite{Adams:2010vc}:
\begin{eqnarray}
S_{\rm inst}(f)&=& \frac{4}{3}\left[ (2 + \cos u)S_{n}^{p}(f) \right. \nonumber \\
&& \left. +4\left(1+ \cos u + \cos^2 u \right)\frac{S_{n}^{a}(f)}{(2\pi f)^{4}}\right] , 
\end{eqnarray}
where $u = 2\pi f/L$, $S_{n}^{p}(f)$ is the position noise and $S_{n}^{a}(f)$ is acceleration noise,
and we have taken the limit of symmetric noise in the detector. The confusion noise due to
gravitational wave sources in the galaxy has been estimated by direct simulation of the LISA
response~\cite{Cornish:2007if} to a synthetic galaxy~\cite{Nelemans:2001hp}, followed by the
removal of resolvable systems~\cite{Crowder:2006eu}.  An approximate fit to the resulting
confusion noise estimate is given by
\begin{equation}
S_{\rm conf}(f) = 
\begin{cases} 
10^{-44.8} f^{-2.4}   &   \, 10^{-4}<f<4.5 \times 10^{-4} \\
10^{-47.15}  f^{-3.1}  &   \, 4.5 \times 10^{-4}<f<1.1 \times 10^{-3}  \\
10^{-51}  f^{-4.4}  &   \, 1.1\times10^{-3}<f<1.7\times10^{-3} \\
10^{-74.7} f^{-13}  &   \, 1.7\times10^{-3}<f<2.5\times10^{-3} \\
10^{-59.15} f^{-7}  &   \, 2.5\times10^{-3}<f<4\times10^{-3}
\end{cases} \, .
\end{equation}
The total noise is then taken to be the sum of the the two contributions: $S_n(f) = S_{\rm inst}(f)
+S_{\rm conf}(f)$.


\section{Parameter Estimation}\label{mcmc}

The goal of a parameter estimation study is to find how accurately we can determine the values of the
source parameters for a given signal. We take a Bayesian approach and use the Markov Chain Monte Carlo
(MCMC) technique to compute the posterior distribution function $p(\vec{x}\vert s)$ describing the
model parameters $\vec{x}$ that we infer from data $s$. Our results establish how well the various
source parameters will be able to be determined by the LISA mission, including the orbital eccentricity.

The use of MCMC techniques in gravitational wave data analysis is now a familiar technique for parameter estimation in gravitational wave astronomy~\cite{Christensen:1998gf,Christensen:2001cr,Cornish:2005qw,Cornish:2006ry,Littenberg:2009bm,vanderSluys:2008qx}.  The result of a well constructed MCMC is a set of samples from the posterior distribution.  The number of samples from a particular region of parameter space is proportional to the posterior weight contained in that region. The uncertainty in each parameter is given by quantiles of the marginalized posterior distribution ({e.g.} the half-width of the $90^{\rm th}$ percentile equates to a 2-$\sigma$ error if the distributions are Gaussian).

By Bayes theorem, the posterior distribution is given by the product of the prior distribution
$p({\bf x})$ and likelihood $p(s \vert {\bf x})$, normalized by the evidence
$p(s)= \int p({\bf x}) p({\bf x}\vert s) \, d{\bf x}$.
For Gaussian noise, the likelihood of the data $s$ having being generated by a gravitational wave signal
$h({\bf x})$ is given by
\begin{equation}\label{likelihood}
p(s \vert {\bf x}) = C e^{-\frac{1}{2}(s-h({\bf x})|s-h({\bf x}))} \ ,
\end{equation}
where $C$ is a normalization constant that does not depend on the signal or the template.
Here we have used the noise weighted inner product
\begin{equation}
(a|b) = 2\int_0^{\infty}\frac{a^{*}(f)b(f) + a(f)b^{*}(f)}{S_n(f)}\,df \ ,
\end{equation}
where $S_n(f)$ is the one-sided noise spectral density.
The prior probability density for ${\bf x}$ reflects our knowledge of the source parameter, however
ill-formed, before we analyze the data. For example, we assume a uniform prior for angular parameters
such as the sky location and the initial orientation of the angular momentum vector, such that
the cosine of the co-latitude is uniformly distributed in the range $[-1:1]$ and the azimuth is
uniformly distributed in the range $[0:2\pi]$.

The primary mode of the posterior distribution yields the best fit values for the source parameters,
according to the current data and our prior knowledge. In many instances the posterior distribution is
multi-modal, and the quantiles used to estimate the parameter uncertainties may cover disjoint regions
in parameter space. Even when the bulk of the posterior weight lies in a single, contiguous region, the
posterior distribution may not be well approximated by a Gaussian distribution. Nonetheless, a
Gaussian approximation to the posterior distribution often provides a reasonable estimate of the
parameter estimation errors, which can be efficiently computed using the Fisher information matrix
$\Gamma_{ij}$, which measures the expectation value of the curvature of the posterior distribution
about the mode:
\begin{equation}
\Gamma_{ij} = -\langle \partial_i \partial_j \ln p({\bf x} \vert s) \rangle\vert_{\rm mode} \, .
\end{equation}

We employ a parallel tempered~\cite{Swendsen:1986} Metropolis-Hastings~\cite{Metropolis:1953ne,Hastings:1970gb}
MCMC routine to explore the PDF.  The Markov chain starts at parameter values ${\bf x}$ and transitions to ${\bf y}$ with probability
\begin{equation}\label{hastings}
H={\rm min}\left[\frac{p({\bf y})p(s \vert {\bf y})q({\bf x}|{\bf y})}{p({\bf x})p(s \vert {\bf x})q({\bf y}|{\bf x})} , 1\right]\ .
\end{equation}
Here $q({\bf x}|{\bf y})$ is the proposal distribution, which is the function that generates proposals for moves from ${\bf x}$ to ${\bf y}$.  The performance of an MCMC algorithm is quite sensitive to the choice of proposal distribution, and care must be taken to ensure that the chains do not get stuck on local maxima of the PDF.  We employed several techniques to ensure rapid exploration of the full parameter space: local coordinate transformations to uncouple the parameters; moves that exploit symmetries of the likelihood surface to encourage jumps between local maxima; and parallel tempering to encourage wide exploration of the posterior~\cite{vanderSluys:2008qx,Key:2008tt,Littenberg:2010gf}.  The number of iterations spent at each parameter value is proportional to the posterior density, and histograms of the parameters visited by the chain provide an estimate of the posterior distribution.

We have found that drawing from a variety of proposal distributions provides a set of jump proposals that tend to produce an MCMC that efficiently maps out the desired PDF and provides accurate parameter uncertainties even for very large search spaces.  Our parameter estimation study thus uses several proposal distributions, including parallel tempering and Fisher matrix proposals.  In this high dimensional parameter space we found it advantageous to use the Fisher matrix approximation to the posterior to propose jumps along single eigen-directions as part of the mixture of jump proposals.

\section{Parameter Estimation with LISA}

We can use our time dependent gravitational waveforms~\cite{Cornish:2010rz} and established MCMC techniques to study how well LISA will be able to measure the full set of seventeen parameters necessary to describe a spinning binary black hole system in an eccentric orbit.  We are especially interested in determining when
eccentric orbits can be distinguished from circular orbits.

We chose to study signals in their final year prior to merger, with an observation time that extends just
beyond the merger.  In order to choose initial parameter values for a system that will merge in one year,
we need to calculate an initial semi-major axis based on the lifetime estimate for a system with some
given initial radial eccentricity. We use the leading order, 0\,PN expression for $a_0$~\cite{Peters:1964en}
\begin{eqnarray}
a_0 = 4\left(\frac{\mu M^2}{5} T_c\right)^{\frac{1}{4}} &&\left(1+\frac{157}{172}e^2_0+\frac{5799977}{7336832}e^4_0 \right. \nonumber\\
&&\left.+\frac{1888175763}{2523870208}e^6_0\right) \, ,
\end{eqnarray}
as an initial guess, and apply a bisection routine to the full numerical orbital evolution to find the value
of $a_0$ that yields a merger time $T_c$ of one year.

\begin{table}[htbp]
    \caption{Parameter ranges for our study of spinning black hole binary systems in eccentric orbits.} 
   \centering
   \begin{tabular}{|c|c|c|} 
      \multicolumn{2}{c}{} \\
      \hline
 \  Parameter  & Minimum   & Maximum  \\ 
	\hline
$m_1$ & $10^5 M_{\odot}$       & $10^7 M_{\odot}$        \\
	\hline
$m_2$ & $10^5 M_{\odot}$       & $ 10^7 M_{\odot}$     \\ 
	\hline
$D_L$ & 1 Gpc        & 100 Gpc        \\
	\hline 
$e_0$ & 0       & 1        \\
	\hline
$a_0$ & 20 M         & 1000 M       \\ 
	\hline
$\chi_1$ & 0         & 1        \\
          \hline
$\chi_2$ & 0       & 1         \\
	\hline
$\cos{\theta}$ & -1           & 1       \\ 
	\hline
$\cos{\theta_L}$ & -1         & 1       \\	
         \hline
$\cos{\theta_{S_1}}$ & -1           & 1       \\
	\hline
$\cos{\theta_{S_2}}$ &-1           & 1      \\ 
	\hline
$\phi$ & 0         & $ 2\pi$        \\
	\hline 
$\phi_L$ & 0         & $ 2\pi$           \\
	\hline
$\phi_{S_1}$ & 0         & $ 2\pi$         \\ 
	\hline
$\phi_{S_2}$ & 0         & $ 2\pi$          \\
          \hline
$n_0$ & 0         & $ 2\pi$         \\
	\hline
$\phi_0$ & 0         & $ 2\pi$         \\ 
	\hline    
      \bottomrule
   \end{tabular}
   \label{tab:ranges}
\end{table}

We use parameter ranges consistent with typical LISA sources, given in Table~\ref{tab:ranges}.  The masses are given in terms of the mass of the Sun, $M_{\odot}=1.9891\times10^{30}$ kg, and the luminosity distance $D_L$ is given in units of Gigaparsecs.  The dimensionless spin parameters $\chi_1$ and $\chi_2$ combine with the black hole masses to give the magnitudes of the spins, $S_i=\chi_i m^2_i$.  There are initial orientation parameters for the orbital angular momentum vector ${\bf L} \rightarrow (\cos{\theta_L},\phi_L)$, as well as the spin vectors ${\bf S}_i\rightarrow (\cos{\theta_{S_i}},\phi_{S_i})$.  The final two parameters, $n_0$ and $\phi_0$, are initial phases related to the mean motion and orbital phase.  In the circular limit these parameters are degenerate, but for eccentric orbits we have to specify the initial periastron position. In the present study we have ignored the possibility that gas dynamics may partially
aligned the spins with the orbital angular momentum~\cite{Bogdanovic:2007hp}, which would restrict the prior ranges and
reduce the degree of orbital precession. The impact of this partial alignment on parameter estimation has
been considered for circular orbits~\cite{Lang:2011je}, and it would be interesting to extend this study to include
eccentricity.

Here we study several representative cases to establish the parameter recovery errors and to study
correlations between the parameters. The high dimension of the parameter space makes it difficult to
perform a comprehensive study -  if we were to choose just two values of each parameter we would need
to perform $2^{17}\sim10^5$ parameter estimation studies. Each MCMC run involves $\sim 100,000$ iterations
with $\sim 8$ parallel chains, and takes about a week to run on a single 2.66 GHz Intel processor, so
we are limited in the number of examples we can consider. We perform MCMC parameter estimation studies of
several representative examples, varying the mass ratio, sky location, distance, eccentricity, and
dimensionless spin parameters.  We only looked at a few initial spin and orbital orientations since
we do not expect these to have a significant effect on the results - unless the initial orientations are very
special the system will explore a wide range of orientations during the orbital evolution. To be able to
explore the parameter space more widely would take a faster code. One possibility is to use the Fisher
information matrix approximation to the posterior, which is many orders of magnitude faster than a
full MCMC study. In preparation for such a study we compare the Fisher matrix approximation to the
MCMC derived posterior distributions and find that the approximation is fairly reliable so
long as the initial eccentricity exceeds $e_0 \sim 0.01$.

\begin{table}[htbp]
   \caption{Injected parameter values for two sets of sources studied with a range of values for $e_0$.  The results of the parameter estimation study for Source 1 are given in Figure~\ref{fig:eccentricity} and the results for Source 2 are given in Figure~\ref{fig:sky1ecc}.}
   \centering
   \begin{tabular}{|c|c|c|} 
      \multicolumn{2}{c}{} \\
      \hline
 \  Parameter  & Source 1 & Source 2  \\ 
	\hline
$m_1$ & $2\times10^6 M_{\odot}$   &    $2\times10^6 M_{\odot}$   \\
	\hline
$m_2$ & $1\times10^6 M_{\odot}$     &  $1\times10^6 M_{\odot}$      \\ 
	\hline
$D_L$ & 6.36167 Gpc  &       6.36167 Gpc     \\
	\hline 
$\chi_1$ & 0.5   &   0.5    \\
          \hline
$\chi_2$ & 0.8 &    0.8  \\
	\hline
$\cos{\theta}$ & 0.2   &  0.4  \\ 
	\hline
$\cos{\theta_L}$ & -0.5  &  -0.5 \\	
         \hline
$\cos{\theta_{S_1}}$ & -0.8    &  -0.8  \\
	\hline
$\cos{\theta_{S_2}}$ & 0.6 & 0.6   \\ 
	\hline
$\phi$ & 1.2  &  2.0  \\
	\hline 
$\phi_L$ & 2.6   &   2.6 \\
	\hline
$\phi_{S_1}$ & 0.4  &  0.4  \\ 
	\hline
$\phi_{S_2}$ & 1.7  &   1.7    \\
          \hline
$n_0$ & 0.2    &  0.2  \\
	\hline
$\phi_0$ &    1.65 &  1.65 \\ 
	\hline    
      \bottomrule
   \end{tabular}
   \label{tab:sourceP}
\end{table}

\begin{figure}[htbp]
   \includegraphics[width=3in]{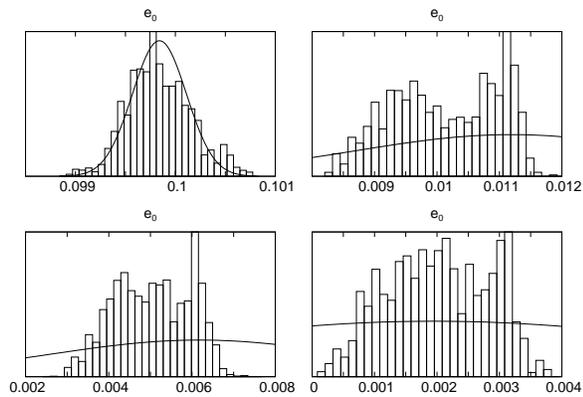}
   \caption{Marginalized posterior distribution for the initial radial eccentricity for sources
 with the same parameter values except for different initial eccentricities and semi-major axes.
The boxed histograms are derived from
the Markov chains, while the solid lines are the Fisher matrix predictions computed
at the MAP values of the parameters (which are pushed off the true values by the instrument noise).
Top left $e_0=0.1$, $a_0=69.96 M$,
SNR$=231$; top right $e_0=0.01$, $a_0=68.33 M$, SNR$=237$; bottom left $e_0=0.005$,
$a_0=69.3 M$, SNR$=237$; bottom right $e_0=0.002$, $a_0=69.3 M$, SNR$=237$. 
The other parameter values correspond to Source 1 in Table~\ref{tab:sourceP}.}
   \label{fig:eccentricity}
\end{figure}

\begin{figure}[htbp]
   \centering
    \begin{tabular}{cc}
      \includegraphics[width=1.6in]{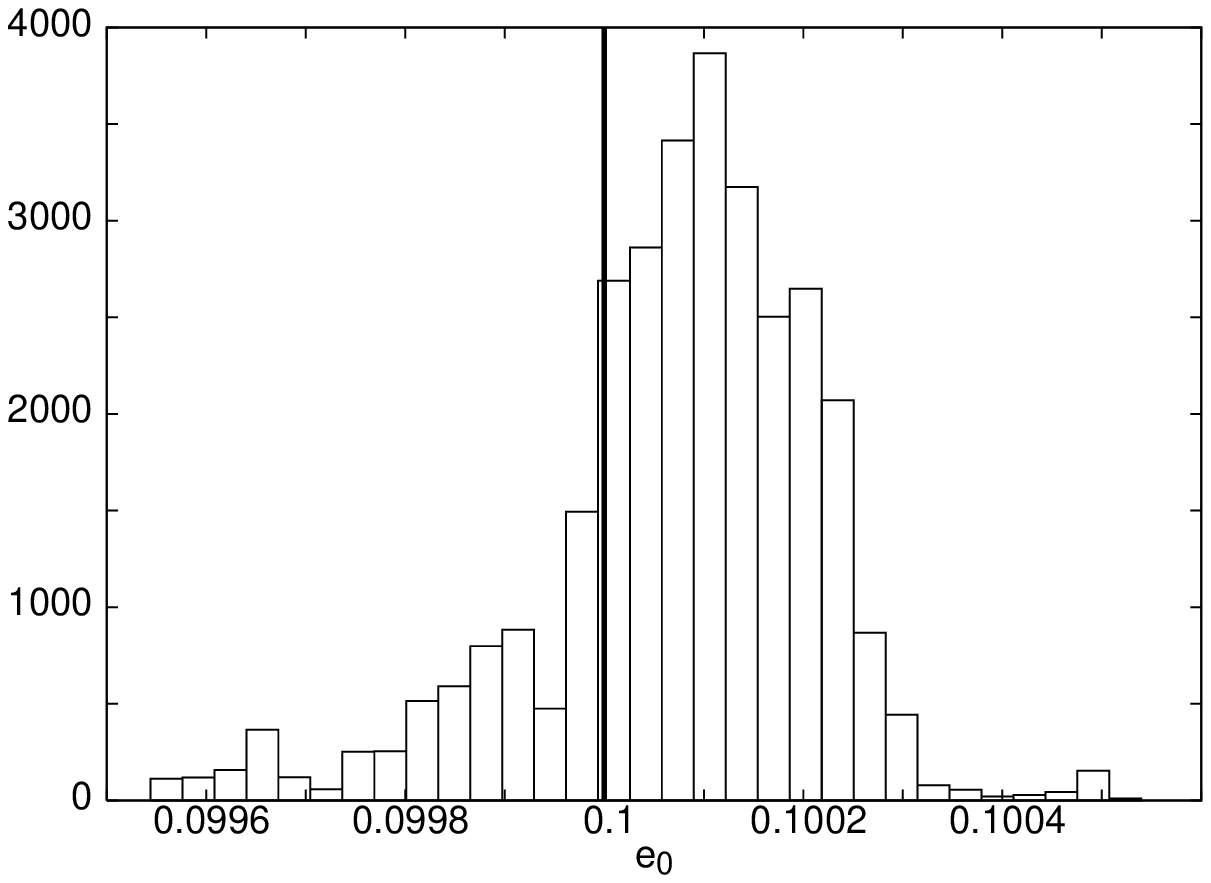} &
      \includegraphics[width=1.6in]{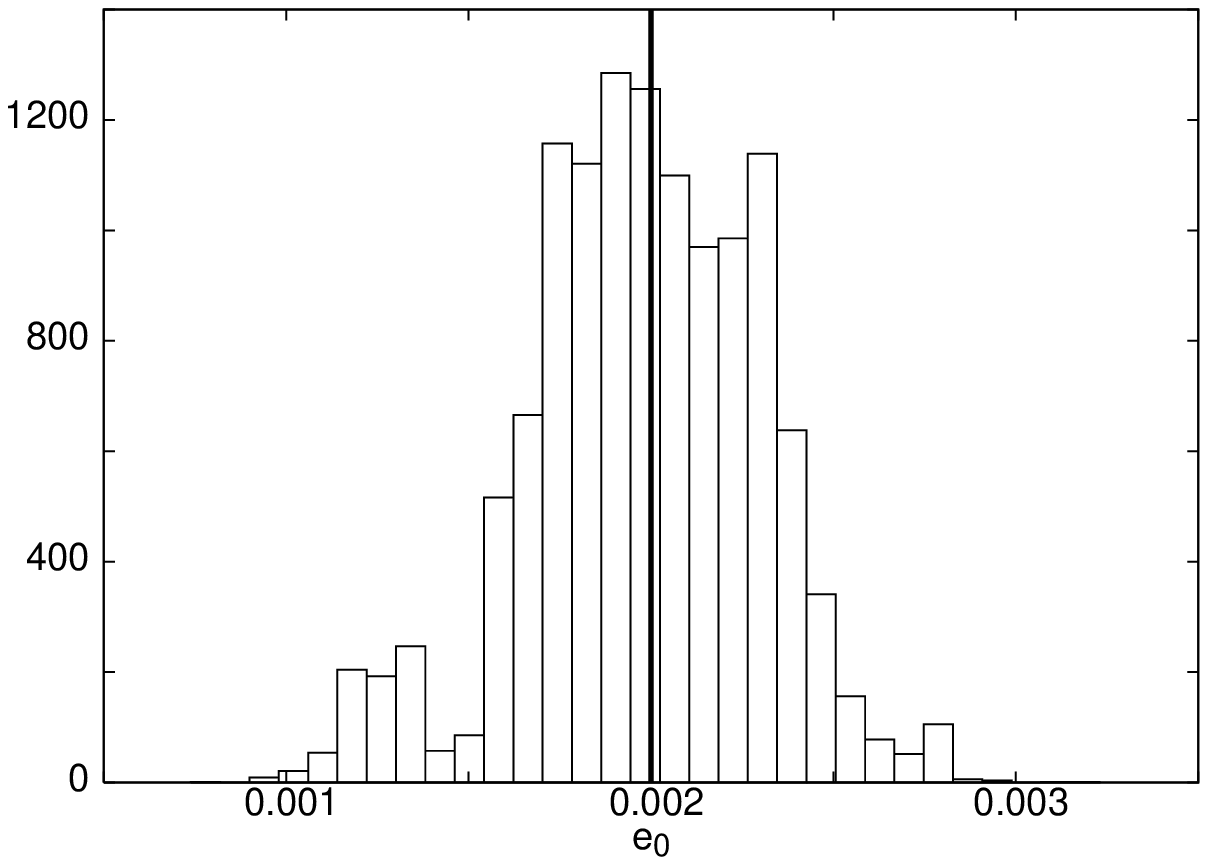} \\
    \end{tabular}
   \caption{The marginalized posterior distribution for the initial radial eccentricity for sources with the same parameter values except for different initial eccentricities and semi-major axes.  On the left $e_0=0.1$, $a_0=68.94 M$, SNR$=557$; on the right $e_0=0.002$, $a_0=68.32 M$, SNR$=559$.  For this source, $e_0=0.002$ is distinguishable from the circular case.  A vertical line marks the injected $e_{0}$ value in each plot.  The other parameter values are given in Table~\ref{tab:sourceP}, Source 2.}
   \label{fig:sky1ecc}
\end{figure}

Our first study focuses on determining when the eccentricity one year before merger is distinguishable
from zero. We studied two systems that only differed in sky location (and hence in SNR),
and considered initial eccentricities in the range $e_0 \in [0.001,0.2]$, see Table~\ref{tab:sourceP}
for a list of the other source parameters.  Marginalized posterior distributions for $e_0$ are shown
in Figure~\ref{fig:eccentricity} for Source 1, and Figure~\ref{fig:sky1ecc} for Source 2. 
We see that the error in the measured value of $e_0$ gets smaller as $e_0$ gets larger, but the
dependence on $e_0$ is fairly weak, and never larger than $\Delta e_0 \sim 0.001$.
Our criteria for deciding if the eccentricity is distinguishable from zero is to see if their is any weight in
the posterior distribution at $e_0=0$ (this test is motivated by the Savage-Dicke density
ratio estimate for the model evidence~\cite{Cornish:2007if}). For Source 1 we see that the examples with
$e_0 \geq 0.005$ are clearly distinguishable from circular, while the $e_0=0.002$ case is on the margin of
detectability. For Source 2, which has higher SNR due to a more favorable sky location, the
$e_0=0.002$ case is clearly distinguishable from circular. The Fisher matrix approximation to the
posterior distribution is computed at the maximum a posteriori probability (MAP) value of the parameters,
and is found to work well for eccentricities $e_0 > 0.05$, but breaks down for small eccentricities.
The MCMC derived posterior distributions are much flatter than a Gaussian distribution, and we attribute
the failure of our Fisher matrix estimates to only including the leading order, quadratic curvature terms
in the Fisher matrix calculation.

This study suggests that LISA observations of eccentric black hole binary systems will be able to measure the eccentricity of the system and distinguish eccentric systems from circular systems to parts in a thousand.  A similar study was performed by Porter and Sesana~\cite{Porter:2010mb} for non-spinning eccentric binary black hole systems.  Their results suggest that LISA will be able to measure the eccentricity to parts in $10^{-4}$ for such sources. 

\begin{figure}[htbp]
   \centering
   \includegraphics[width=3in]{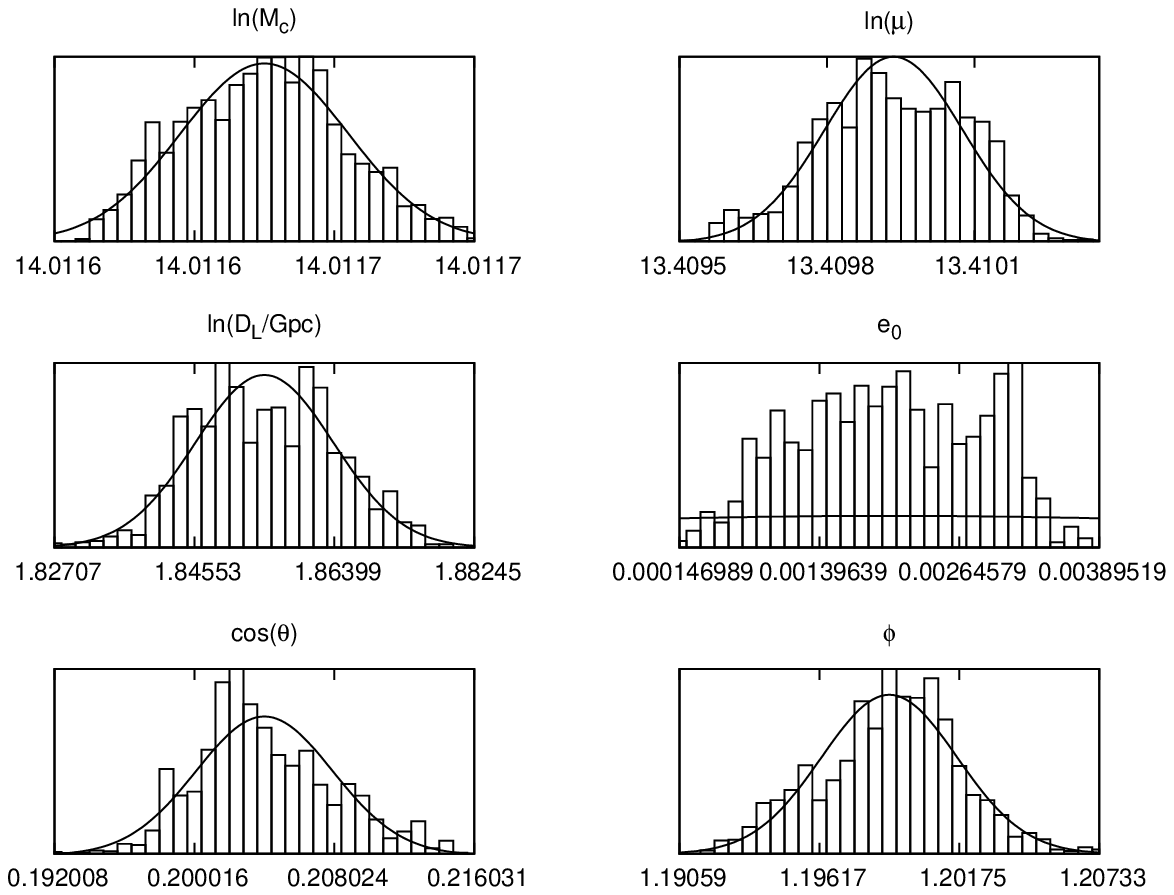}
   \caption{The marginalized posterior distribution for several source parameters, including the initial radial eccentricity with injected value $e_0=0.002$, $a_0=69.3 M$, and SNR$=237$. The other injected source parameters
are those of Source 1 in Table~\ref{tab:sourceP}.  The solid lines are the Fisher matrix predictions computed
at the MAP values of the parameters.}
   \label{fig:ecc0.3}
\end{figure}

\begin{figure}[htbp]
   \centering
   \includegraphics[width=3in]{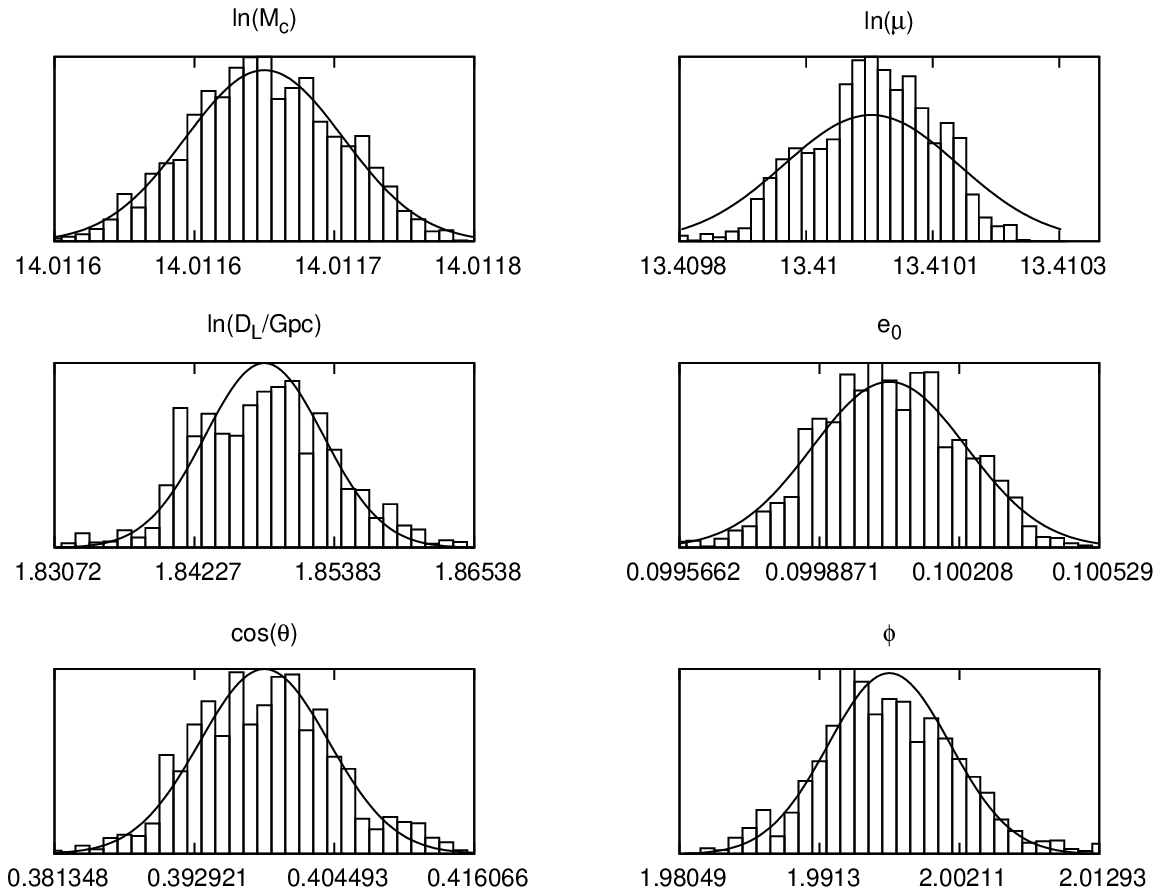}
   \caption{The marginalized posterior distribution for several source parameters, including the initial radial eccentricity with injected value $e_0=0.1$, $a_0=68.94 M$, and SNR $=558$. The other injected source parameters
are those of Source 2 in Table~\ref{tab:sourceP}.  The solid lines are the Fisher matrix predictions computed
at the MAP values of the parameters.}
   \label{fig:sky1_0.1}
\end{figure}

We find that the other source parameters are also measured quite well, as illustrated in
Figure~\ref{fig:ecc0.3} and Figure~\ref{fig:sky1_0.1}.  Marginalized posterior distributions are shown for the 
chirp mass $M_c= (m_1 m_2)^{3/5}/(m_1+m_2)^{1/5}$ and reduced mass $\mu = m_1 m_2 /(m_1 + m_2)$,
the distance to the source, the initial radial eccentricity, and the two sky location parameters.
We compare the Fisher matrix approximation to the marginalized posterior distributions computed from the
MCMC runs and find excellent agreement for all the parameters (except for the eccentricity in Figure~\ref{fig:ecc0.3}).

As a start to exploring the large parameter space of eccentric binary black hole systems we consider a few representative examples below.  We find that in general physical parameters are well constrained by LISA observations and that the Fisher matrix makes fair estimates of the parameter errors.  In addition to the examples below with varied spin and distance parameters we also studied systems with a range of mass ratios $m_1/m_2 \in [1,5]$ and total masses $M \in [10^5 M_{\odot},10^7 M_{\odot}]$ with similar results. We also compared the parameter estimation errors with those obtained when
the eccentricity is held fixed at zero, and saw only small (less than 50\%) changes in the error estimates.

\begin{table}[htbp]
   \caption{Source 3 gives the injected parameter values for Figure~\ref{fig:hist2} with large spin values.  Source 4 gives the injected parameter values for Figure~\ref{fig:hist1} with small spin values.}
   \centering
   \begin{tabular}{|c|c|c|} 
      \multicolumn{2}{c}{} \\
      \hline
 \  Parameter  & Source 3 & Source 4  \\ 
	\hline
$m_1$ & $2\times10^6 M_{\odot}$   &    $2\times10^6 M_{\odot}$   \\
	\hline
$m_2$ & $1\times10^6 M_{\odot}$     &  $1\times10^6 M_{\odot}$      \\ 
	\hline
$D_L$ & 6.36167 Gpc  &       6.36167 Gpc     \\
	\hline 
$e_0$ & 0.1     &    0.1     \\
	\hline
$a_0$ &   68.96 M &  68.91 M  \\ 
	\hline
$\chi_1$ & 0.8   &   0.1    \\
          \hline
$\chi_2$ & 0.9 &    0.11  \\
	\hline
$\cos{\theta}$ & 0.2   &  0.2  \\ 
	\hline
$\cos{\theta_L}$ & -0.5  &  -0.5 \\	
         \hline
$\cos{\theta_{S_1}}$ & -0.8    &  -0.8  \\
	\hline
$\cos{\theta_{S_2}}$ & 0.6 & 0.6   \\ 
	\hline
$\phi$ & 1.2  &  1.2  \\
	\hline 
$\phi_L$ & 2.6   &   2.6 \\
	\hline
$\phi_{S_1}$ & 0.4  &  0.4  \\ 
	\hline
$\phi_{S_2}$ & 1.7  &   1.7    \\
          \hline
$n_0$ & 0.2    &  0.2  \\
	\hline
$\phi_0$ &    1.65 &  1.65 \\ 
	\hline    
	\hline
SNR & 250 & 197 \\	
	\hline    
      \bottomrule
   \end{tabular}
   \label{tab:sourceP2}
\end{table}

\begin{figure}[htbp]
   \centering
   \includegraphics[width=3in]{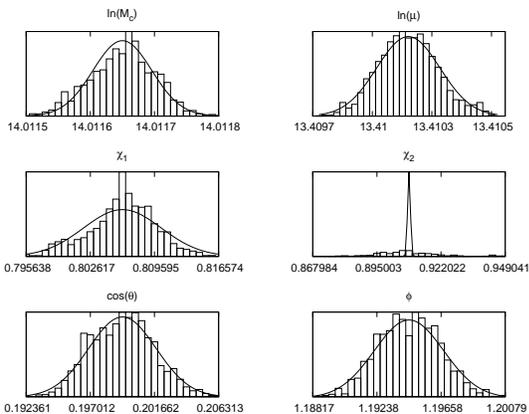}
   \caption{The marginalized posterior distribution for several source parameters for
Source 3 in Table~\ref{tab:sourceP2}.
The solid lines are the Fisher matrix predictions computed at the MAP values of the parameters.}
   \label{fig:hist2}
\end{figure}

\begin{figure}[htbp]
   \centering
   \includegraphics[width=3in]{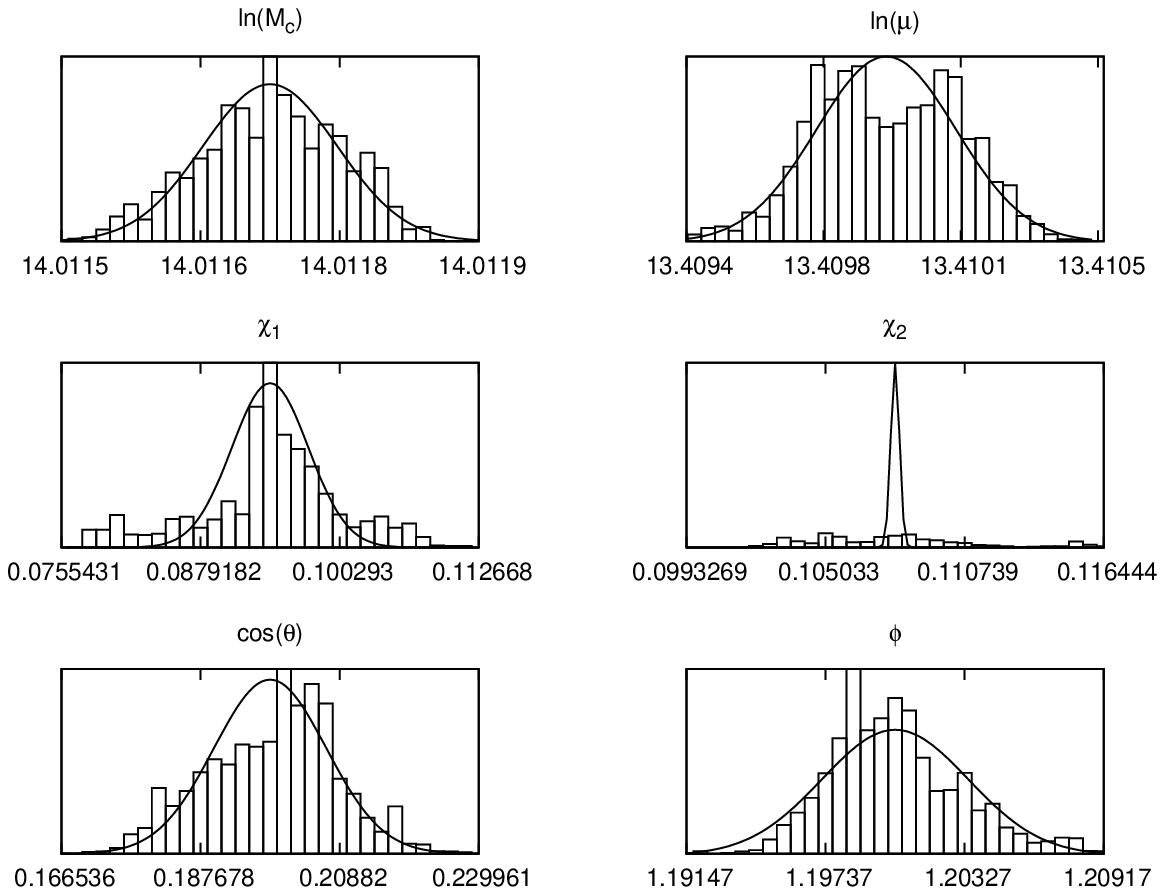}
   \caption{The marginalized posterior distribution for several source parameters for
Source 4 in Table~\ref{tab:sourceP2}.
The solid lines are the Fisher matrix predictions computed at the MAP values of the parameters.}
   \label{fig:hist1}
\end{figure}

\begin{figure}[htbp]
   \centering
   \includegraphics[width=3in]{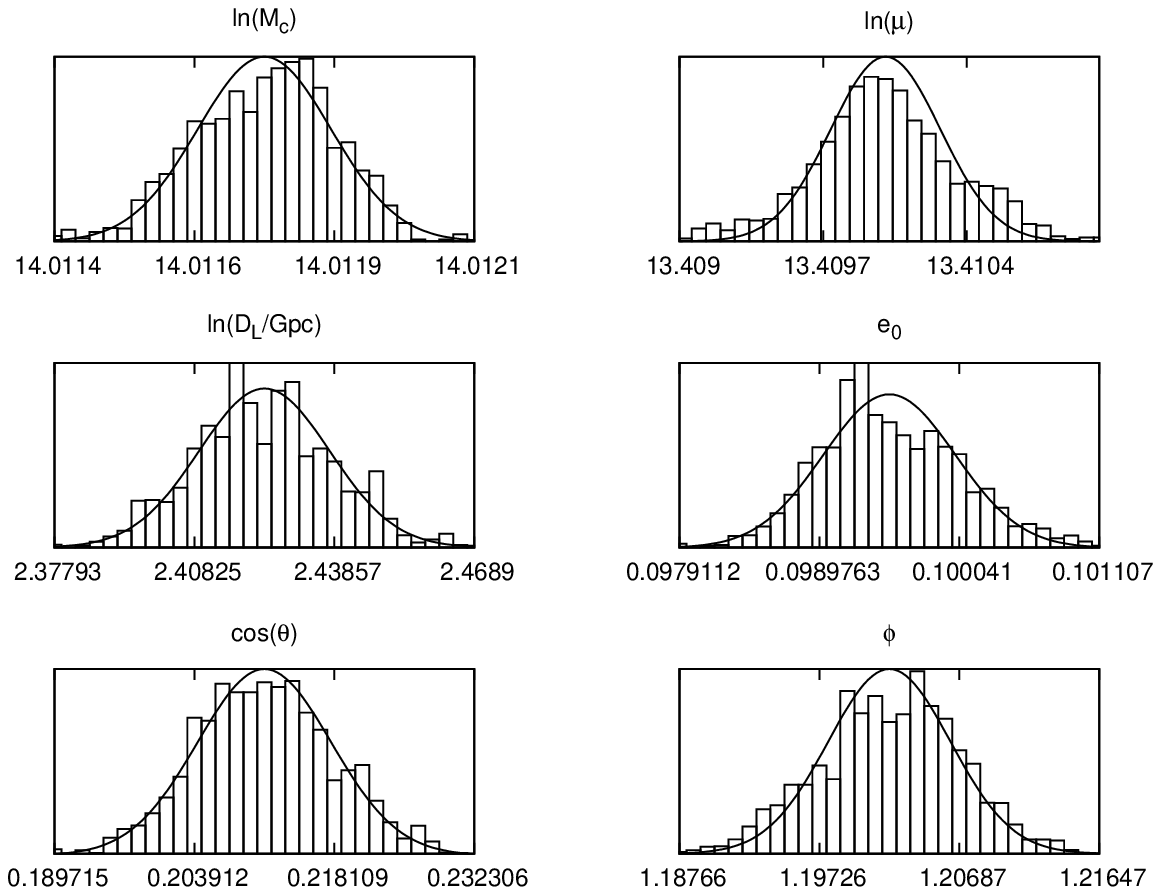}
   \caption{The marginalized posterior distribution for several parameters for a source with $z \sim 1.5$ $e_{0}=0.1$, and SNR $=132$ (Table~\ref{tab:sourceP3}, Source 5).
The solid lines are the Fisher matrix predictions computed at the MAP values of the parameters.
}
   \label{fig:dist1}
\end{figure}

\begin{figure}[htbp]
   \centering
   \includegraphics[width=3in]{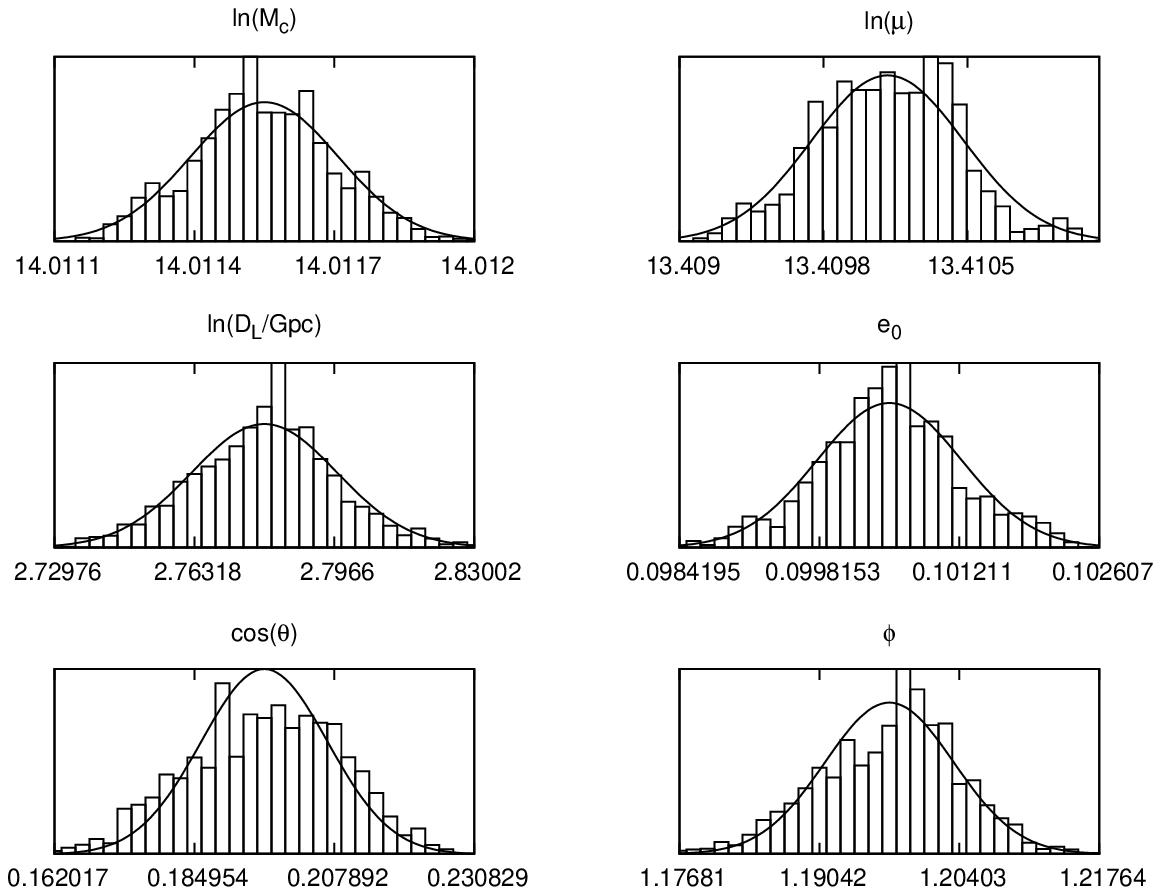}
   \caption{The marginalized posterior distribution for several parameters for a source with $z\sim 2$, $e_{0}=0.1$, and SNR $=92$ (Table~\ref{tab:sourceP3}, Source 6). The solid lines are the Fisher matrix predictions
computed at the MAP values of the parameters.}
   \label{fig:dist2}
\end{figure}

The dimensionless spin parameters $\chi_1$ and $\chi_2$ are varied in Table~\ref{tab:sourceP2} to study the cases of
large and small spin parameters.  The magnitude of the spin vectors is related to the mass of the black
hole $S_i=\chi_i m^2_i$. The posterior distribution for several of the source parameters for these cases are
given in Figures~\ref{fig:hist2} and~\ref{fig:hist1}.  The Fisher approximation is a reasonable prediction
of the width of the posterior distribution for the spin of the more massive body, but does a poor job for the
less massive body. This discrepancy was seen in many other examples that we looked at. The cause of the
discrepancy is presently not understood. Two lower SNR examples are shown in Figure~\ref{fig:dist1} and
Figure~\ref{fig:dist2}, and we see good agreement with the Fisher matrix estimates.

\begin{table}[htbp]
   \caption{Source 5 gives the injected parameter values for Figure~\ref{fig:dist1} with redshift $z\sim1.5$.  Source 6 gives the injected parameter values for Figure~\ref{fig:dist2} with redshift $z\sim2$.}
   \centering
   \begin{tabular}{|c|c|c|} 
      \multicolumn{2}{c}{} \\
      \hline
 \  Parameter  & Source 5 & Source 6  \\ 
	\hline
$m_1$ & $2\times10^6 M_{\odot}$   &    $2\times10^6 M_{\odot}$   \\
	\hline
$m_2$ & $1\times10^6 M_{\odot}$     &  $1\times10^6 M_{\odot}$      \\ 
	\hline
$D_L$ & 11.008 Gpc  &      15.733  Gpc     \\
	\hline 
$e_0$ & 0.1     &    0.1     \\
	\hline
$a_0$ &   68.94 M &  68.94 M  \\ 
	\hline
$\chi_1$ & 0.5   &   0.8    \\
          \hline
$\chi_2$ & 0.5 &    0.8  \\
	\hline
$\cos{\theta}$ & 0.2   &  0.2  \\ 
	\hline
$\cos{\theta_L}$ & -0.5  &  -0.5 \\	
         \hline
$\cos{\theta_{S_1}}$ & -0.8    &  -0.8  \\
	\hline
$\cos{\theta_{S_2}}$ & 0.6 & 0.6   \\ 
	\hline
$\phi$ & 1.2  &  1.2  \\
	\hline 
$\phi_L$ & 2.6   &   2.6 \\
	\hline
$\phi_{S_1}$ & 0.4  &  0.4  \\ 
	\hline
$\phi_{S_2}$ & 1.7  &   1.7    \\
          \hline
$n_0$ & 0.2    &  0.2  \\
	\hline
$\phi_0$ &    1.65 &  1.65 \\ 
	\hline    
	\hline
SNR & 132 & 92 \\	
	\hline    
      \bottomrule
   \end{tabular}
   \label{tab:sourceP3}
\end{table}

\begin{figure}[htbp]
   \centering
       \includegraphics[width=3in]{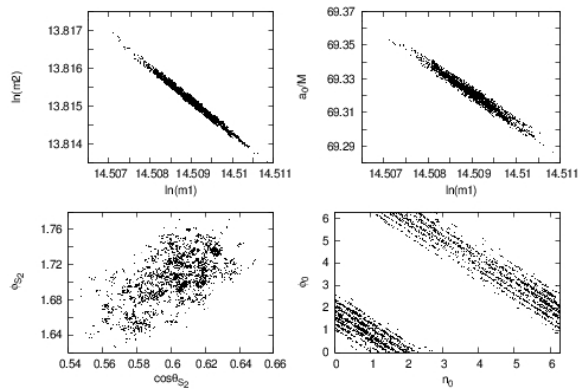}
  \caption{Two dimensional posterior distribution scatter plots showing the correlation between pairs of parameters for a source with initial radial eccentricity $e_0=0.002$ and SNR $=237$ (Table~\ref{tab:sourceP}, Source 1).  Upper left $m_1$ and $m_2$, upper right $m_1$ and $a_0$, bottom left $\cos \theta_{S_2}$ and $\phi_{S_2}$, bottom right $n_0$ and $\phi_0$.}
   \label{fig:masses}   
\end{figure}

\begin{figure}[htbp]
   \centering
      \includegraphics[width=3in]{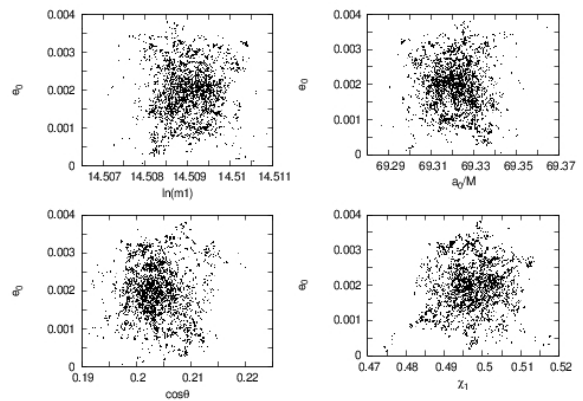} 
   \caption{Two dimensional posterior distribution scatter plots for a source with initial radial eccentricity $e_0=0.002$ and SNR $=237$ (Table~\ref{tab:sourceP}, Source 1).  Upper left $m_1$ and $e_0$, upper right $a_0$ and $e_0$, bottom left $\cos \theta$ and $e_0$, bottom right $\chi_1$ and $e_0$.}
   \label{fig:ecchist}   
\end{figure}

We do not expect the eccentricity to be highly correlated with the other source parameters since the higher harmonics
introduced in the waveform due to eccentricity cannot be simulated by changes in other parameters or their
combinations.  We indeed find that the initial eccentricity is not correlated with the other parameters.  Compare
the distribution of values for the two masses in Figure~\ref{fig:masses} to the distribution of mass and eccentricity
values in Figure~\ref{fig:ecchist}.  The two masses are highly correlated since it is the total mass of the system
and the ratio of the masses that appear in the waveform.  The distribution of eccentricity versus the other source
parameters is similar to that seen in Figure~\ref{fig:ecchist}.

\section{Conclusion}\label{con}

Our studies of the response of the LISA detector to the gravitational wave signal from spinning
binary black hole systems in eccentric orbits show that the eccentricity should not be neglected for
LISA data analysis and parameter estimation.  We find that LISA can determine the eccentricity of the
system one year before merger to parts in a thousand.  This result depends only weakly on the initial
value of the eccentricity, indicating that LISA will be able to distinguish between eccentric and
circular orbits at the same level $\left(e_0\sim10^{-3}\right)$. 

The construction of the gravitational waveforms for spinning binary black hole systems in eccentric orbits to 1.5\,PN order in~\cite{Cornish:2010rz} establishes the framework for the extension of this work to higher post-Newtonian order.  Binary black hole waveforms that include eccentricity will be necessary for several of the LISA science goals including constraining galaxy merger scenarios and testing general relativity in the strong field regime near supermassive black holes.

This work builds the foundation for further studies including a comprehensive exploration of the parameter space including source sky location, distance, spins, masses, and mass ratios.  For sources with initial radial eccentricity greater than $e_0 \sim 0.02$ the Fisher matrix is a good approximation to the posterior distribution.  The very large parameter space could be studied quickly using the Fisher approximation, although it is not as useful for the low initial eccentricity cases.

The equations of motion and instantaneous gravitational waveforms have been calculated to the next post-Newtonian order and these pieces can be included in a future parameter estimation study.  The 2\,PN effects include the spin-spin coupling of the two black holes and thus corrections to the precession and evolution of the system.  

These waveforms can be also be used to study how well the Advanced LIGO-Virgo network and proposed Einstein Telescope will be able to measure eccentricity and what level of bias could be expected from using circular templates for parameter estimation.  


\section{Acknowledgements}	
  
This work was supported by NASA grants NNX07AJ61G and NNX10AH15G. We are grateful to Alberto Sesana and Edward Porter
for extensive and informative discussion of the mechanisms responsible for producing black hole binaries with
eccentric orbits.


\begin{thebibliography}{99}

\expandafter\ifx\csname bibnamefont\endcsname\relax
  \def\bibnamefont#1{#1}\fi
\expandafter\ifx\csname bibfnamefont\endcsname\relax
  \def\bibfnamefont#1{#1}\fi
\expandafter\ifx\csname citenamefont\endcsname\relax
  \def\citenamefont#1{#1}\fi

\bibitem[{\citenamefont{Micic et~al.}(2007)\citenamefont{Micic,
  Holley-Bockelmann, Sigurdsson, and Abel}}]{Micic:2007vd}
\bibinfo{author}{\bibfnamefont{M.}~\bibnamefont{Micic}},
  \bibinfo{author}{\bibfnamefont{K.}~\bibnamefont{Holley-Bockelmann}},
  \bibinfo{author}{\bibfnamefont{S.}~\bibnamefont{Sigurdsson}},
  \bibnamefont{and} \bibinfo{author}{\bibfnamefont{T.}~\bibnamefont{Abel}},
  \bibinfo{journal}{Mon. Not. Roy. Astron. Soc.}
  \textbf{\bibinfo{volume}{380}}, \bibinfo{pages}{1533} (\bibinfo{year}{2007}),
  \eprint{astro-ph/0703540}.

\bibitem[{\citenamefont{Sesana et~al.}(2007)\citenamefont{Sesana, Volonteri,
  and Haardt}}]{Sesana:2007sh}
\bibinfo{author}{\bibfnamefont{A.}~\bibnamefont{Sesana}},
  \bibinfo{author}{\bibfnamefont{M.}~\bibnamefont{Volonteri}},
  \bibnamefont{and} \bibinfo{author}{\bibfnamefont{F.}~\bibnamefont{Haardt}},
  \bibinfo{journal}{Mon. Not. Roy. Astron. Soc.}
  \textbf{\bibinfo{volume}{377}}, \bibinfo{pages}{1711} (\bibinfo{year}{2007}),
  \eprint{astro-ph/0701556}.

\bibitem[{\citenamefont{Kormendy and Richstone}(1995)}]{Kormendy:1995er}
\bibinfo{author}{\bibfnamefont{J.}~\bibnamefont{Kormendy}} \bibnamefont{and}
  \bibinfo{author}{\bibfnamefont{D.}~\bibnamefont{Richstone}},
  \bibinfo{journal}{Ann. Rev. Astron. Astrophys.}
  \textbf{\bibinfo{volume}{33}}, \bibinfo{pages}{581} (\bibinfo{year}{1995}).

\bibitem[{\citenamefont{Magorrian et~al.}(1998)}]{Magorrian:1997hw}
\bibinfo{author}{\bibfnamefont{J.}~\bibnamefont{Magorrian}}
  \bibnamefont{et~al.}, \bibinfo{journal}{Astron. J.}
  \textbf{\bibinfo{volume}{115}}, \bibinfo{pages}{2285} (\bibinfo{year}{1998}),
  \eprint{astro-ph/9708072}.

\bibitem[{\citenamefont{Narayan}(2005)}]{Narayan:2005oq}
\bibinfo{author}{\bibfnamefont{R.}~\bibnamefont{Narayan}},
  \bibinfo{journal}{New Journal of Physics} \textbf{\bibinfo{volume}{7}},
  \bibinfo{pages}{199} (\bibinfo{year}{2005}),

\bibitem[{\citenamefont{Begelman et~al.}(1980)\citenamefont{Begelman,
  Blandford, and Rees}}]{Begelman:1980vb}
\bibinfo{author}{\bibfnamefont{M.~C.} \bibnamefont{Begelman}},
  \bibinfo{author}{\bibfnamefont{R.~D.} \bibnamefont{Blandford}},
  \bibnamefont{and} \bibinfo{author}{\bibfnamefont{M.~J.} \bibnamefont{Rees}},
  \bibinfo{journal}{Nature} \textbf{\bibinfo{volume}{287}},
  \bibinfo{pages}{307} (\bibinfo{year}{1980}).

\bibitem{Peters:63}
  P.~C.~Peters and J.~Mathews,
  Phys.\ Rev.\ {\bf 131}, 435 (1963).

\bibitem[{\citenamefont{{Peters}}(1964)}]{Peters:1964en}
\bibinfo{author}{\bibfnamefont{P.~C.} \bibnamefont{{Peters}}},
  \bibinfo{journal}{Physical Review} \textbf{\bibinfo{volume}{136}},
  \bibinfo{pages}{1224} (\bibinfo{year}{1964}).


\bibitem[{\citenamefont{Krolak and Schutz}(1987)}]{Krolak:1987vp}
\bibinfo{author}{\bibfnamefont{A.}~\bibnamefont{Krolak}} \bibnamefont{and}
  \bibinfo{author}{\bibfnamefont{B.~F.} \bibnamefont{Schutz}},
  \bibinfo{journal}{General Relativity and Gravitation}
  \textbf{\bibinfo{volume}{19}}, \bibinfo{pages}{1163} (\bibinfo{year}{1987}).

\bibitem[{\citenamefont{Colpi et~al.}(1999)\citenamefont{Colpi, Mayer, and
  Governato}}]{Colpi:1999cm}
\bibinfo{author}{\bibfnamefont{M.}~\bibnamefont{Colpi}},
  \bibinfo{author}{\bibfnamefont{L.}~\bibnamefont{Mayer}}, \bibnamefont{and}
  \bibinfo{author}{\bibfnamefont{F.}~\bibnamefont{Governato}},
  \bibinfo{journal}{arXiv} \textbf{\bibinfo{volume}{astro-ph/9907088}}
  (\bibinfo{year}{1999}), \eprint{astro-ph/9907088}.

\bibitem[{\citenamefont{Gultekin et~al.}(2006)\citenamefont{Gultekin,
  Coleman~Miller, and Hamilton}}]{Gultekin:2005fd}
\bibinfo{author}{\bibfnamefont{K.}~\bibnamefont{Gultekin}},
  \bibinfo{author}{\bibfnamefont{M.}~\bibnamefont{Coleman~Miller}},
  \bibnamefont{and} \bibinfo{author}{\bibfnamefont{D.~P.}
  \bibnamefont{Hamilton}}, \bibinfo{journal}{Astrophys. J.}
  \textbf{\bibinfo{volume}{640}}, \bibinfo{pages}{156} (\bibinfo{year}{2006}),
  \eprint{astro-ph/0509885}.

\bibitem[{\citenamefont{Dotti et~al.}(2006)\citenamefont{Dotti, Colpi, and
  Haardt}}]{Dotti:2006ae}
\bibinfo{author}{\bibfnamefont{M.}~\bibnamefont{Dotti}},
  \bibinfo{author}{\bibfnamefont{M.}~\bibnamefont{Colpi}}, \bibnamefont{and}
  \bibinfo{author}{\bibfnamefont{F.}~\bibnamefont{Haardt}},
  \bibinfo{journal}{arXiv} \textbf{\bibinfo{volume}{astro-ph/0602013}}
  (\bibinfo{year}{2006}), \eprint{astro-ph/0602013}.

\bibitem[{\citenamefont{Berczik et~al.}(2005)\citenamefont{Berczik, Merritt,
  and Spurzem}}]{Berczik:2005ff}
\bibinfo{author}{\bibfnamefont{P.}~\bibnamefont{Berczik}},
  \bibinfo{author}{\bibfnamefont{D.}~\bibnamefont{Merritt}}, \bibnamefont{and}
  \bibinfo{author}{\bibfnamefont{R.}~\bibnamefont{Spurzem}},
  \bibinfo{journal}{Astrophys. J.} \textbf{\bibinfo{volume}{633}},
  \bibinfo{pages}{680} (\bibinfo{year}{2005}), \eprint{astro-ph/0507260}.

\bibitem[{\citenamefont{Armitage and Natarajan}(2005)}]{Armitage:2005xq}
\bibinfo{author}{\bibfnamefont{P.~J.} \bibnamefont{Armitage}} \bibnamefont{and}
  \bibinfo{author}{\bibfnamefont{P.}~\bibnamefont{Natarajan}},
  \bibinfo{journal}{Astrophys. J.} \textbf{\bibinfo{volume}{634}},
  \bibinfo{pages}{921} (\bibinfo{year}{2005}), \eprint{astro-ph/0508493}.

\bibitem[{\citenamefont{Aarseth}(2003)}]{Aarseth:2002ie}
\bibinfo{author}{\bibfnamefont{S.~J.} \bibnamefont{Aarseth}},
  \bibinfo{journal}{Astrophys. Space Sci.} \textbf{\bibinfo{volume}{285}},
  \bibinfo{pages}{367} (\bibinfo{year}{2003}), \eprint{astro-ph/0210116}.



\bibitem{AmaroSeoane:2006py}
  P.~Amaro-Seoane and M.~Freitag,
  Astrophys.\ J.\  {\bf 653}, L53 (2006)
  [arXiv:astro-ph/0610478].

\bibitem{AmaroSeoane:2007aw}
  P.~Amaro-Seoane, J.~R.~Gair, M.~Freitag, M.~Coleman Miller, I.~Mandel, C.~J.~Cutler and S.~Babak,
  Class.\ Quant.\ Grav.\  {\bf 24}, R113 (2007)
  [arXiv:astro-ph/0703495].

\bibitem{AmaroSeoane:2009yr}
  P.~Amaro-Seoane, C.~Miller and M.~Freitag,
  Astrophys.\ J.\  {\bf 692}, L50 (2009)
  [arXiv:0901.0604 [astro-ph.SR]].

\bibitem{AmaroSeoane:2009cg}
  P.~Amaro-Seoane, C.~Eichhorn, E.~Porter and R.~Spurzem,
  arXiv:0908.0755 [astro-ph.GA].

\bibitem{Sesana:2010qb}
  A.~Sesana,
  arXiv:1006.0730 [astro-ph.CO]

\bibitem{Plowman:2009rp}
  J.~E.~Plowman, D.~C.~Jacobs, R.~W.~Hellings, S.~L.~Larson and S.~Tsuruta,
 ``Constraining the Black Hole Mass Spectrum with Gravitational Wave Observations I: The Error Kernel"
  arXiv:0903.2059 [astro-ph].

\bibitem{Sesana:2010wy}
  A.~Sesana, J.~R.~Gair, E.~Berti and M.~Volonteri,
  arXiv:1011.5893 [astro-ph.CO].

\bibitem{Plowman:2010fc}
  J.~E.~Plowman, R.~W.~Hellings and S.~Tsuruta,
  arXiv:1009.0765 [astro-ph.CO].



\bibitem[{\citenamefont{Martel and Poisson}(1999)}]{Martel:1999kb}
\bibinfo{author}{\bibfnamefont{K.}~\bibnamefont{Martel}} \bibnamefont{and}
  \bibinfo{author}{\bibfnamefont{E.}~\bibnamefont{Poisson}},
  \bibinfo{journal}{Physical Review D} \textbf{\bibinfo{volume}{60}},
  \bibinfo{pages}{124008} (\bibinfo{year}{1999}),

\bibitem[{\citenamefont{Brown and Zimmerman}(2010)}]{Brown:2009ng}
\bibinfo{author}{\bibfnamefont{D.~A.} \bibnamefont{Brown}} \bibnamefont{and}
  \bibinfo{author}{\bibfnamefont{P.~J.} \bibnamefont{Zimmerman}},
  \bibinfo{journal}{Phys. Rev.} \textbf{\bibinfo{volume}{D81}},
  \bibinfo{pages}{024007} (\bibinfo{year}{2010}), \eprint{0909.0066}.

\bibitem[{\citenamefont{Cutler and Vallisneri}(2007)}]{Cutler:2007mi}
\bibinfo{author}{\bibfnamefont{C.}~\bibnamefont{Cutler}} \bibnamefont{and}
  \bibinfo{author}{\bibfnamefont{M.}~\bibnamefont{Vallisneri}},
  \bibinfo{journal}{Phys. Rev.} \textbf{\bibinfo{volume}{D76}},
  \bibinfo{pages}{104018} (\bibinfo{year}{2007}), \eprint{0707.2982}.

\bibitem{Porter:2010mb}
  E.~K.~Porter and A.~Sesana,
  ``Eccentric Massive Black Hole Binaries in LISA I : The Detection Capabilities of Circular Templates,''
  arXiv:1005.5296 [gr-qc].

\bibitem[{\citenamefont{Kidder}(1995)}]{Kidder:1995zr}
\bibinfo{author}{\bibfnamefont{L.~E.} \bibnamefont{Kidder}},
  \bibinfo{journal}{Phys. Rev.} \textbf{\bibinfo{volume}{D52}},
  \bibinfo{pages}{821} (\bibinfo{year}{1995}), \eprint{gr-qc/9506022}.

\bibitem{Faye:2006gx}
  G.~Faye, L.~Blanchet and A.~Buonanno,
  ``Higher-order spin effects in the dynamics of compact binaries. I: Equations of motion,''
  Phys.\ Rev.\  D {\bf 74}, 104033 (2006)
  
\bibitem{Blanchet:2006gy}
  L.~Blanchet, A.~Buonanno and G.~Faye,
  ``Higher-order spin effects in the dynamics of compact binaries II. Radiation field,''
  Phys.\ Rev.\  D {\bf 74}, 104034 (2006)

\bibitem[{\citenamefont{Majar and Vasuth}(2008)}]{Majar:2008zz}
\bibinfo{author}{\bibfnamefont{J.}~\bibnamefont{Majar}} \bibnamefont{and}
  \bibinfo{author}{\bibfnamefont{M.}~\bibnamefont{Vasuth}},
  \bibinfo{journal}{Phys. Rev.} \textbf{\bibinfo{volume}{D77}},
  \bibinfo{pages}{104005} (\bibinfo{year}{2008}), \eprint{0806.2273}.

\bibitem{Yunes:2009ke}
      Nicolas Yunes and Frans Pretorius
       ``Fundamental Theoretical Bias in Gravitational Wave Astrophysics and the Parameterized Post-Einsteinian Framework"
      Phys.\ Rev.\ D {\bf 80}, 122003 (2009)

\bibitem[{\citenamefont{Wen}(2003)}]{Wen:2003ec}
\bibinfo{author}{\bibfnamefont{L.}~\bibnamefont{Wen}}, \bibinfo{journal}{The
  Astrophysical Journal} \textbf{\bibinfo{volume}{598}}, \bibinfo{pages}{419}
  (\bibinfo{year}{2003})

\bibitem[{\citenamefont{O'Leary et~al.}(2009)\citenamefont{O'Leary, Kocsis, and
  Loeb}}]{OLeary:2008kx}
\bibinfo{author}{\bibfnamefont{R.~M.} \bibnamefont{O'Leary}},
  \bibinfo{author}{\bibfnamefont{B.}~\bibnamefont{Kocsis}}, \bibnamefont{and}
  \bibinfo{author}{\bibfnamefont{A.}~\bibnamefont{Loeb}},
  \bibinfo{journal}{MNRAS}  (\bibinfo{year}{2009}),


\bibitem{Cornish:2010rz}
  N.~J.~Cornish and J.~S.~Key,
  Phys.\ Rev.\  D {\bf 82}, 044028 (2010)
  arXiv:1004.5322 [gr-qc].

\bibitem{Lang:1900bz}
  R.~N.~Lang and S.~A.~Hughes,
  Phys.\ Rev.\  D {\bf 74}, 122001 (2006)
  [Erratum-ibid.\  D {\bf 75}, 089902 (2007\ ERRAT,D77,109901.2008)]
  [arXiv:gr-qc/0608062].

\bibitem{Arun:2008zn}
  K.~G.~Arun {\it et al.},
  Class.\ Quant.\ Grav.\  {\bf 26}, 094027 (2009)
  [arXiv:0811.1011 [gr-qc]].



\bibitem[{\citenamefont{Estabrook et~al.}(2000)\citenamefont{Estabrook, Tinto,
  and Armstrong}}]{Estabrook:2000ef}
\bibinfo{author}{\bibfnamefont{F.~B.} \bibnamefont{Estabrook}},
  \bibinfo{author}{\bibfnamefont{M.}~\bibnamefont{Tinto}}, \bibnamefont{and}
  \bibinfo{author}{\bibfnamefont{J.~W.} \bibnamefont{Armstrong}},
  \bibinfo{journal}{Phys. Rev. D} \textbf{\bibinfo{volume}{62}}
  (\bibinfo{year}{2000}).

\bibitem[{\citenamefont{Prince et~al.}(2002)\citenamefont{Prince, Tinto,
  Larson, and Armstrong}}]{Prince:2002bx}
\bibinfo{author}{\bibfnamefont{T.~A.} \bibnamefont{Prince}},
  \bibinfo{author}{\bibfnamefont{M.}~\bibnamefont{Tinto}},
  \bibinfo{author}{\bibfnamefont{S.~L.} \bibnamefont{Larson}},
  \bibnamefont{and}
  \bibinfo{author}{\bibfnamefont{J.}~\bibnamefont{Armstrong}},
  \bibinfo{journal}{Phys. Rev. D} \textbf{\bibinfo{volume}{66}}
  (\bibinfo{year}{2002}).

\bibitem{Cornish:2001bb}
  N.~J.~Cornish,
  ``Detecting a stochastic gravitational wave background with the laser interferometer space antenna,''
  Phys.\ Rev.\  D {\bf 65}, 022004 (2002)

\bibitem{Adams:2010vc}
  M.~R.~Adams and N.~J.~Cornish,
  ``Discriminating between a Stochastic Gravitational Wave Background and Instrument Noise,''
  arXiv:1002.1291 [gr-qc].
  
\bibitem{Nelemans:2001hp}
  G.~Nelemans, L.~R.~Yungelson and S.~F.~Portegies Zwart,
  ``The gravitational wave signal from the galactic disk population of binaries containing two compact objects,''
 A\&A 375 3  890-898 (2001)

\bibitem{Cornish:2007if}
  N.~J.~Cornish and T.~B.~Littenberg,
  Phys.\ Rev.\  D {\bf 76}, 083006 (2007)

\bibitem{Crowder:2006eu}
  J.~Crowder and N.~Cornish,
  Phys.\ Rev.\  D {\bf 75}, 043008 (2007)


\bibitem{Christensen:1998gf}
  N.~Christensen and R.~Meyer,
  ``Markov chain Monte Carlo methods for Bayesian gravitational radiation data analysis,''
  Phys.\ Rev.\  D {\bf 58}, 082001 (1998)

\bibitem{Christensen:2001cr}
  N.~Christensen and R.~Meyer,
  ``Using Markov chain Monte Carlo methods for estimating parameters with gravitational radiation data,''
  Phys.\ Rev.\  D {\bf 64}, 022001 (2001)

\bibitem{Cornish:2005qw}
  N.~J.~Cornish and J.~Crowder,
  Phys.\ Rev.\  D {\bf 72}, 043005 (2005)
  [arXiv:gr-qc/0506059].

\bibitem[{\citenamefont{Cornish and Porter}(2006)}]{Cornish:2006ry}
\bibinfo{author}{\bibfnamefont{N.~J.} \bibnamefont{Cornish}} \bibnamefont{and}
  \bibinfo{author}{\bibfnamefont{E.~K.} \bibnamefont{Porter}},
  \bibinfo{journal}{Class. Quant. Grav.} \textbf{\bibinfo{volume}{23}},
  \bibinfo{pages}{S761} (\bibinfo{year}{2006}), \eprint{gr-qc/0605085}.

\bibitem{Littenberg:2009bm}
  T.~B.~Littenberg and N.~J.~Cornish,
  Phys.\ Rev.\  D {\bf 80}, 063007 (2009)
  [arXiv:0902.0368 [gr-qc]].

\bibitem[{\citenamefont{van der Sluys~et. al.}(2008)}]{vanderSluys:2008qx}
\bibinfo{author}{\bibfnamefont{M.}~\bibnamefont{van der Sluys~et. al.}},
  \bibinfo{journal}{Class. Quant. Grav.} \textbf{\bibinfo{volume}{25}},
  \bibinfo{pages}{184011} (\bibinfo{year}{2008}), \eprint{0805.1689}.

\bibitem[{\citenamefont{Key and Cornish}(2009)}]{Key:2008tt}
\bibinfo{author}{\bibfnamefont{J.~S.} \bibnamefont{Key}} \bibnamefont{and}
  \bibinfo{author}{\bibfnamefont{N.~J.} \bibnamefont{Cornish}},
  \bibinfo{journal}{Phys. Rev.} \textbf{\bibinfo{volume}{D79}},
  \bibinfo{pages}{043014} (\bibinfo{year}{2009}), \eprint{0812.1590}.

 \bibitem{Swendsen:1986}
R.H. Swendsen and J.S. Wang
Phys. Rev. Lett. 57, 2607 (1986). 

\bibitem[{\citenamefont{Metropolis et~al.}(1953)\citenamefont{Metropolis,
  Rosenbluth, Rosenbluth, Teller, and Teller}}]{Metropolis:1953ne}
\bibinfo{author}{\bibfnamefont{N.}~\bibnamefont{Metropolis}},
  \bibinfo{author}{\bibfnamefont{A.}~\bibnamefont{Rosenbluth}},
  \bibinfo{author}{\bibfnamefont{M.}~\bibnamefont{Rosenbluth}},
  \bibinfo{author}{\bibfnamefont{A.}~\bibnamefont{Teller}}, \bibnamefont{and}
  \bibinfo{author}{\bibfnamefont{E.}~\bibnamefont{Teller}},
  \bibinfo{journal}{Journal of Chemical Physics} \textbf{\bibinfo{volume}{21}},
  \bibinfo{pages}{1087} (\bibinfo{year}{1953}).

\bibitem[{\citenamefont{Hastings}(1970)}]{Hastings:1970gb}
\bibinfo{author}{\bibfnamefont{W.}~\bibnamefont{Hastings}},
  \bibinfo{journal}{Biometrika} \textbf{\bibinfo{volume}{97}},
  \bibinfo{pages}{57} (\bibinfo{year}{1970}).

\bibitem[{\citenamefont{Cornish and Crowder}(2005)}]{Cornish:2005ul}
\bibinfo{author}{\bibfnamefont{N.~J.} \bibnamefont{Cornish}} \bibnamefont{and}
  \bibinfo{author}{\bibfnamefont{J.}~\bibnamefont{Crowder}},
  \bibinfo{journal}{Phys. Rev. D} \textbf{\bibinfo{volume}{72}}
  (\bibinfo{year}{2005}).

\bibitem{Littenberg:2010gf}
  T.~B.~Littenberg and N.~J.~Cornish,
  Phys.\ Rev.\  D {\bf 82}, 103007 (2010)
  [arXiv:1008.1577 [gr-qc]].

\bibitem{Bogdanovic:2007hp}
  T.~Bogdanovic, C.~S.~Reynolds and M.~C.~Miller,
  arXiv:astro-ph/0703054.

\bibitem{Lang:2011je}
  R.~N.~Lang, S.~A.~Hughes and N.~J.~Cornish,
  arXiv:1101.3591 [gr-qc].

\bibitem{Ryan:1995wh}
  F.~D.~Ryan,
  ``Gravitational waves from the inspiral of a compact object into a massive,
  axisymmetric body with arbitrary multipole moments,''
  Phys.\ Rev.\  D {\bf 52}, 5707 (1995).

\bibitem{Collins:2004ex}
  N.~A.~Collins and S.~A.~Hughes,
  ``Towards a formalism for mapping the spacetimes of massive compact objects:
  Bumpy black holes and their orbits,''
  Phys.\ Rev.\  D {\bf 69}, 124022 (2004)

\bibitem{Berti:2005ys}
  E.~Berti, V.~Cardoso and C.~M.~Will,
  ``On gravitational-wave spectroscopy of massive black holes with the  space
  interferometer LISA,''
  Phys.\ Rev.\  D {\bf 73}, 064030 (2006)

\bibitem{Berti:2007zu}
  E.~Berti, J.~Cardoso, V.~Cardoso and M.~Cavaglia,
  ``Matched-filtering and parameter estimation of ringdown waveforms,''
  Phys.\ Rev.\  D {\bf 76}, 104044 (2007)

\bibitem{Hughes:2006pm}
  S.~A.~Hughes,
  ``(Sort of) Testing relativity with extreme mass ratio inspirals,''
  AIP Conf.\ Proc.\  {\bf 873}, 233 (2006)

\bibitem{Arun:2006yw}
  K.~G.~Arun, B.~R.~Iyer, M.~S.~S.~Qusailah and B.~S.~Sathyaprakash,
  ``Testing post-Newtonian theory with gravitational wave observations,''
  Class.\ Quant.\ Grav.\  {\bf 23}, L37 (2006)

\bibitem{Arun:2006hn}
  K.~G.~Arun, B.~R.~Iyer, M.~S.~S.~Qusailah and B.~S.~Sathyaprakash,
  ``Probing the non-linear structure of general relativity with black hole
  mergers,''
  Phys.\ Rev.\  D {\bf 74}, 024006 (2006)
  
\end{thebibliography}
\end{document}